\documentclass[twocolumn,showpacs,preprintnumbers,amsmath,amssymb]{revtex4}
\usepackage{dcolumn}

\usepackage{graphicx,amssymb,amsmath}

\begin{document}

\title{Incomplete equilibrium in 
  long-range interacting systems}

\author{Fulvio Baldovin and Enzo Orlandini}
\email{baldovin@pd.infn.it, orlandin@pd.infn.it}
\affiliation{
Dipartimento di Fisica and
Sezione INFN, Universit\`a di Padova,\\
\it Via Marzolo 8, I-35131 Padova, Italy
}

\date{\today}

\begin{abstract}
We use a Hamiltonian dynamics to discuss the statistical
mechanics of long-lasting quasi-stationary states 
particularly relevant for long-range interacting systems.  
Despite the presence of an anomalous single-particle velocity
distribution, we find that the Central Limit Theorem implies the
Boltzmann expression in Gibbs' $\Gamma$-space.  
We identify the nonequilibrium sub-manifold of $\Gamma$-space
characterizing the anomalous behavior and show that by restricting
the Boltzmann-Gibbs approach to this sub-manifold we obtain the statistical
mechanics of the quasi-stationary states. 
\end{abstract}

\maketitle

In comparison with its equilibrium counterpart, 
nonequilibrium statistical mechanics does not rely on universal notions,
like the ensembles ones, through which one can handle large classes of
physical systems \cite{maes}. 
Incomplete (or partial) equilibrium states
\cite{landau,incomplete} are in this respect a remarkable exception, since
in these cases concepts of equilibrium statistical mechanics can be
used to describe nonequilibrium situations.    
Incomplete equilibrium states arise when 
different parts of the system themselves reach a state of equilibrium
long before they equilibrate with each other \cite{landau}.
The classical understanding on how a system approaches equilibrium is
based on the short time-scale collisions mechanism which links any initial
condition to the statistical equilibrium. 
For long-range interacting systems, this picture is not valid anymore
since the time-scale for microscopic collisions diverges with 
the range of the interactions.
This implies
that the Boltzmann equation must be substituted with other
approximations such as the Vlasov or the Balescu-Lenard equations
\cite{balescu}, where the interparticle correlations are negligible or
almost negligible and a nonequilibrium initial configuration  
could stay frozen or almost frozen for a very long time.  
This applies, e.g., to gravitational systems,
Bose-Einstein condensates and plasma physics \cite{dauxois}. 
Due to the physical relevance of long-range interacting systems and to
the privileged position of incomplete equilibrium states in 
nonequilibrium statistical mechanics, 
it is important to investigate whether the notion of incomplete
equilibrium plays an important role in understanding the
nonequilibrium properties of these systems.

Recently we showed \cite{ham_can} that nonequilibrium states in which
the value of macroscopic quantities remains stationary or
quasi-stationary for reasonably long time (quasi-stationary states --
QSSs) are 
important, e.g., for experiments, since they appear even when the
long-range system exchanges energy with a thermal bath (TB). 
Using the same paradigmatic long-range interacting system of
Ref. \cite{ham_can}, the Hamiltonian Mean Field (HMF) model
\cite{konishi}, here we discuss the Gibbs' $\Gamma$-space
statistical mechanics description of the QSSs in a 
canonical ensemble perspective. 
We identify the nonequilibrium sub-manifold of $\Gamma$-space within
which the quasi-stationary dynamics is confined and we show that the
Boltzmann-Gibbs (BG) approach, restricted to this sub-manifold, gives
the correct statistics of the QSSs. 
In this respect, the QSSs can be interpreted as incomplete equilibrium
states \cite{landau}. 
Our theoretical framework
allows one to calculate, on the basis of the empirical
detection of the temperature and of the value of an order parameter,
any other thermodynamic quantity such as the energy or the specific
heat of the system. 
The possibility of predicting physical quantities which characterize 
the QSSs could be useful, i.e., for understanding nonequilibrium
features of gravitational or plasma structures 
and it is then of particular interest for
experimentalists or theorists of long-range interacting systems.
Since the system considered is naturally endowed with a microscopic
Hamiltonian dynamics, we validate step by step our theoretical
derivation  with {\it a priori} results obtained from dynamical
simulations.  
Our findings also furnish novel significant arguments to an intense
debate in the literature \cite{latora,debate_1,debate_2,debate_3},
that so far has been restricted to the
single-particle $\mu$-space and to the microcanonical ensemble.

The HMF model can be introduced as a  
set of $M$  globally coupled $X Y$-spins
with Hamiltonian \cite{konishi} 
\begin{equation}
H_{\rm HMF}=\sum_{i=1}^M\frac{l_i^2}{2}
+\frac{1}{2M}\sum_{i,j=1}^M\left[1-\cos(\theta_i-\theta_j)\right],
\label{hmf}
\end{equation}
where $\theta_i\in[0,2\pi)$ are the spin angles and $l_i\in\mathbb R$ their angular momenta
(velocities). 
  The specific magnetization of the system is 
$m_{\rm HMF}\equiv |\sum_{i=1}^M(\cos\theta_i,\sin\theta_i)|/M$
and the temperature $T$ is identified with twice the specific
kinetic energy.
We have thus 
$e_{\rm HMF}=T_{\rm HMF}/2+(1-m_{\rm HMF}^2)/2$,
where $e_{\rm HMF}\equiv E_{\rm HMF}/M$ is the specific energy. 
Direct connections with the problem of disk
galaxies \cite{chavanis} and free electron lasers experiments \cite{barre}
have been established for this Hamiltonian. Eq. (\ref{hmf}) has also
been shown to be representative of the class of
Hamiltonians on a one-dimensional lattice in which the potential
is proportional to
$\sum_{i,j=1}^M\left[1-\cos(\theta_i-\theta_j)\right]/r_{i
j}^\alpha$, where $r_{i j}$ is the lattice separation between
spins and $\alpha<1$ \cite{tamarit}.
Hence, the Hamiltonian in Eq. (\ref{hmf}) can be considered as an
interesting ``paradigm'' for long-range interacting systems
\cite{chavanis}. 
The TB introduced in \cite{ham_can} is characterized by
$N\gg M$ equivalent spins first-neighbors coupled along a
chain
\begin{equation}
H_{\rm TB}=\sum_{i=M+1}^{N}\frac{l_i^2}{2}
+\sum_{i=M+1}^{N}\left[1-\cos(\theta_{i+1}-\theta_i)\right],
\label{tb}
\end{equation}
with $\theta_{N+1}\equiv\theta_{M+1}$, and 
the interaction between HMF and TB is 
given by
\begin{equation}
H_{\rm I}=\epsilon\sum_{i=1}^{M}\sum_{s=1}^S\left[1-\cos(\theta_{i}-\theta_{r_s(i)})\right],
\label{interaction}
\end{equation}
where $\epsilon$ is a coupling constant that modulates the interaction
strength between HMF and TB.
Each HMF-spin is thus in contact with $S$ 
TB-spins specified as initial
condition ($r_s(i)$ are independent integer random numbers in the interval
$[M+1,N]$). 
A ``surface-like effect'' $S\sim M^{\gamma-1}$ ($0<\gamma<1$)
guarantees a consistent thermodynamic limit \cite{ham_can}.
For $\epsilon=0$ HMF and TB are decoupled and the setup 
reproduces the microcanonical dynamics. 
For $\epsilon \ne 0$ the whole
system is at constant energy 
whereas the energy of the HMF model fluctuates. 
Our numerics are obtained with  
$M=10^3$, $N=M^2$, $S=10^5M^{-1/2}$, $0.005\leq\epsilon\leq0.1$ (we
use dimensionless units),
through a velocity-Verlet algorithm 
assuring a total energy conservation within an error  
$\Delta E/E<10^{-5}$ \cite{ham_can}.  
The width $T_0$ of the Maxwellian probability density function 
(PDF) for the initial TB-velocities 
is a control parameter for the bath temperature. 
For $\epsilon>0$ we showed \cite{ham_can} that 
the HMF temperature finally converges to the BG
equilibrium at temperature $T_0$. 

\begin{figure}
\includegraphics[width=0.49\columnwidth]{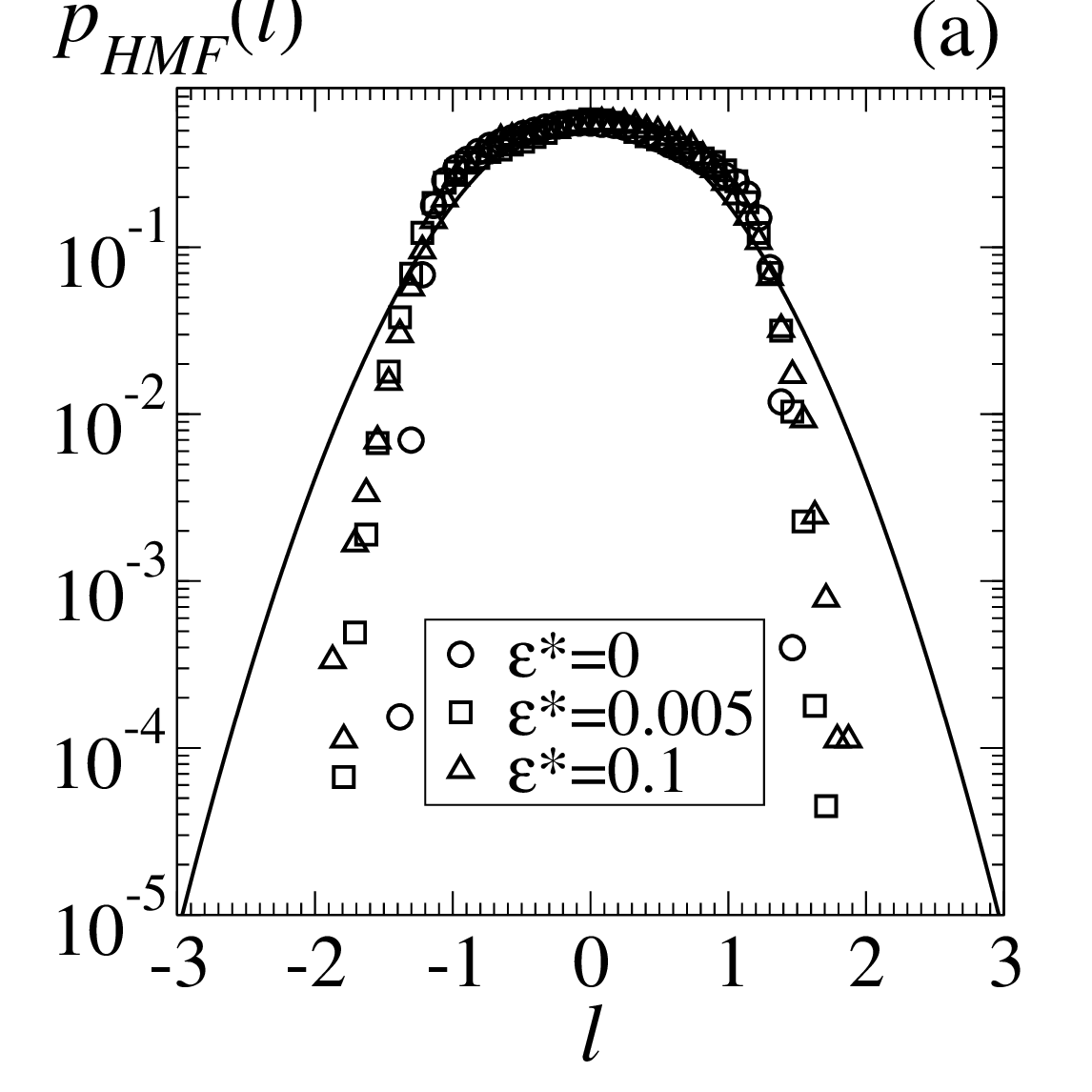}
\includegraphics[width=0.49\columnwidth]{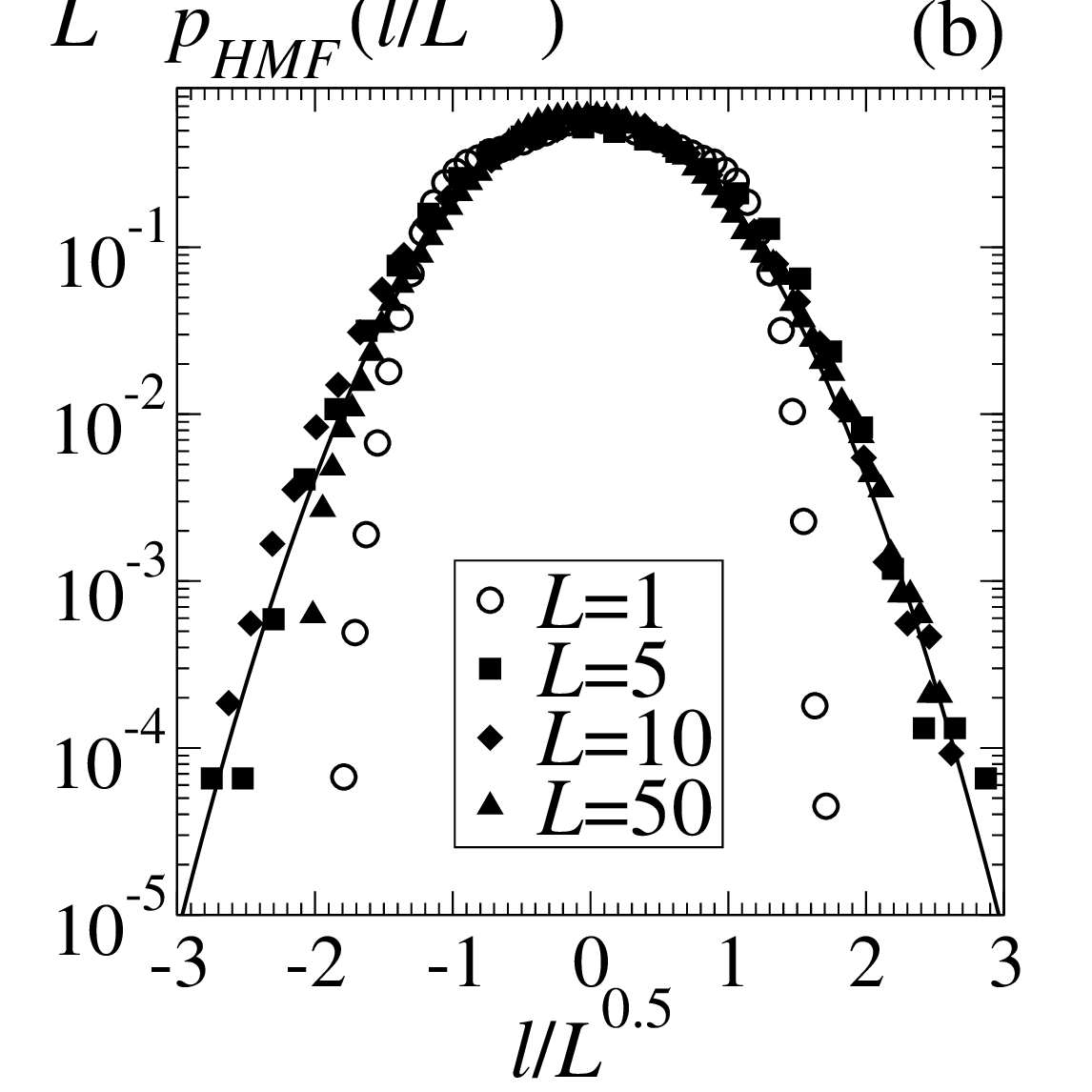}\\
\vspace{0.25cm}
\includegraphics[width=0.98\columnwidth]{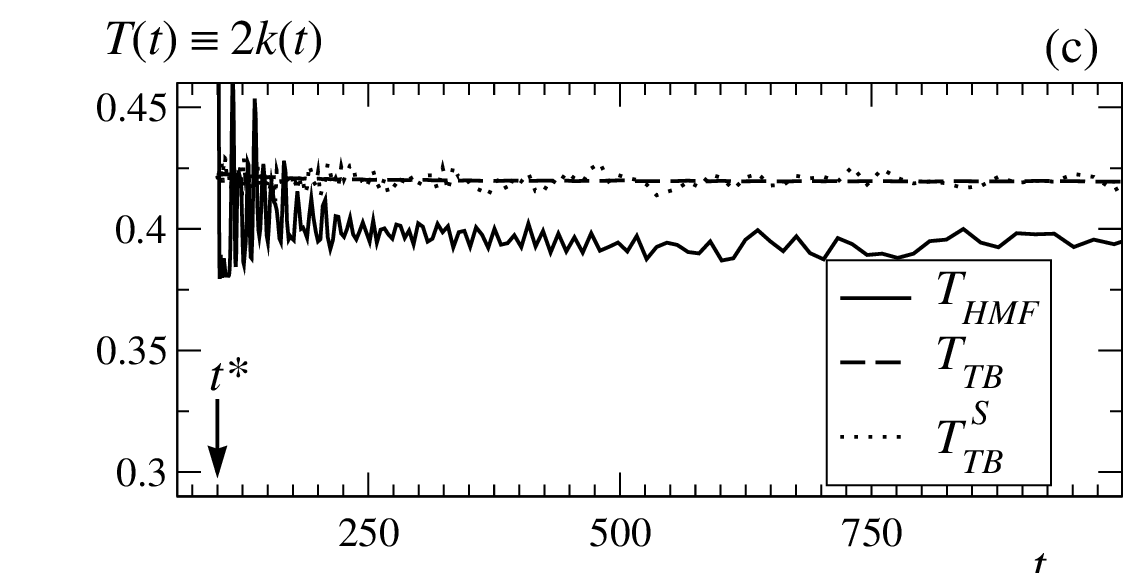}
\caption{
QSS for $M=10^3$ and $T_0=0.38$. For $\epsilon=0.005$ we observe the
average values $m_{\rm HMF}^2\simeq 0.015$ and $T_{\rm HMF}\simeq 0.397$ 
($e_{\rm HMF}\simeq 0.691$).
(a): Single particle velocity PDF. The solid line is $p_{\rm TB}(l)$.
(b): PDF of the sum of the velocities of $L$ particles. By multiplying
the PDFs for different $L$'s by $L^{0.5}$ and dividing
the velocities $l$ by $L^{0.5}$, all data collapse fairly well
on to the line that corresponds to a Gaussian distribution 
with width $T=0.397$. 
(c): Temperature time-evolutions for three different subsets of the
system during the QSSs. Here the TB temperature has been shifted to
$T_0=0.42$.
}
\label{fig_qss_1}
\end{figure}

By setting far-from-equilibrium initial
conditions for the HMF model,
the relaxation to equilibrium
typically displays stationary or quasi-stationary stages during which
the phase functions $m_{\rm HMF}$, $T_{\rm HMF}$ (and thus also $e_{\rm HMF}$)
fluctuate around constant or almost constant average values~\cite{ham_can}. 
This behavior is particularly interesting when the life-time of the
QSS diverges in the thermodynamic limit  
\cite{ham_can,latora,debate_1,debate_2,debate_3}. 
This happens if for
example at $t=0$ we set a delta distribution for the angles 
($p_{\rm HMF}(\theta)=\delta(0)\Rightarrow
m_{\rm HMF}^2=1$), a uniform distribution for the velocities,
$p_{\rm HMF}(l)=1/2\bar l,\;l\in[-\bar l,\bar l]$, 
with $\bar l\simeq2.03$ ($e_{\rm HMF}\simeq0.69$) \cite{ham_can},  
and a TB temperature  $T_0=0.38$. 
In Fig. \ref{fig_qss_1}a we show that during the QSS, for
$\epsilon>0$, the single particle velocity PDF is non-Maxwellian 
and similar to the distribution found in the microcanonical case
\cite{latora,debate_1,debate_2} ($\epsilon=0$).  

Given some probability distribution for  
the initial data, a dynamical estimation of phase functions, like
e.g. the energy $E_{\rm HMF}$, can be obtained 
by recording the phase function 
values at different times in a single orbit
and averaging over different realizations of the initial conditions. 
To understand the connection between the
anomalous PDF in $\mu$-space and the $\Gamma$-space statistics, 
we start by measuring the PDF of the
{\it sum} of the velocities of $L$ particles, $p_{\rm HMF}^L$
(Fig. \ref{fig_qss_1}b).   
Such a distribution 
tends very quickly to the Gaussian form as $L$ increases. 
In fact, a rescaling of
$l$ by $L^{1/2}$ and a multiplication of $p_{\rm HMF}^L(l)$ by the same
factor reveal the Central Limit Theorem (CLT) data collapse onto the
Maxwellian (Gaussian) distribution of temperature $T=T_{\rm HMF}=0.397$. 
The fact that the CLT applies to the sum of the velocities 
is a strong indication \cite{kinchin} that in $\Gamma$-space 
the probability for the energy $E_{\rm HMF}$ is characterized by the
Boltzmann expression 
$\omega(E_{\rm HMF})e^{-E_{\rm HMF}/T}$
($k_B\equiv1$), where $\omega(E_{\rm HMF})$ is a
density of states.
Although this situation resembles equilibrium, there are some important
differences. 
For example, 
the anomalous velocity PDF in $\mu$-space
implies that the joint probability of all particles is not
given by a mere product of exponentials. 
The Boltzmann expression arises, because of weak enough
particle-particle correlations \cite{balescu}, for a sufficiently
large number of particles. 
Below, we directly verify its occurrence. 

Another key observation is that during the QSS the HMF 
does not thermalize with the TB. 
In Fig. \ref{fig_qss_1}c we shifted the TB temperature by $10\%$, 
setting it to $T_0=0.42$. While this modifies the final HMF
equilibrium temperature, it does not affect $T_{\rm HMF}$ during the QSS.  
Even  the subset of $S$ TB-spins 
in direct contact with the HMF model, 
$\{\theta_{r_s(i)}\}_{1\leq s\leq S,1\leq i\leq M}$,  is at 
$T_{\rm TB}^S=T_0$ and does not thermalize with the HMF temperature. 
The energy fluctuations are nevertheless significantly larger than those due
to the algorithm precision 
($\Delta E_{\rm HMF}/E_{\rm HMF}\simeq 10^{-2}$ for $M=10^3$),
distinguishing the canonical QSSs from the microcanonical ones.

\begin{figure}
\includegraphics[width=0.98\columnwidth]{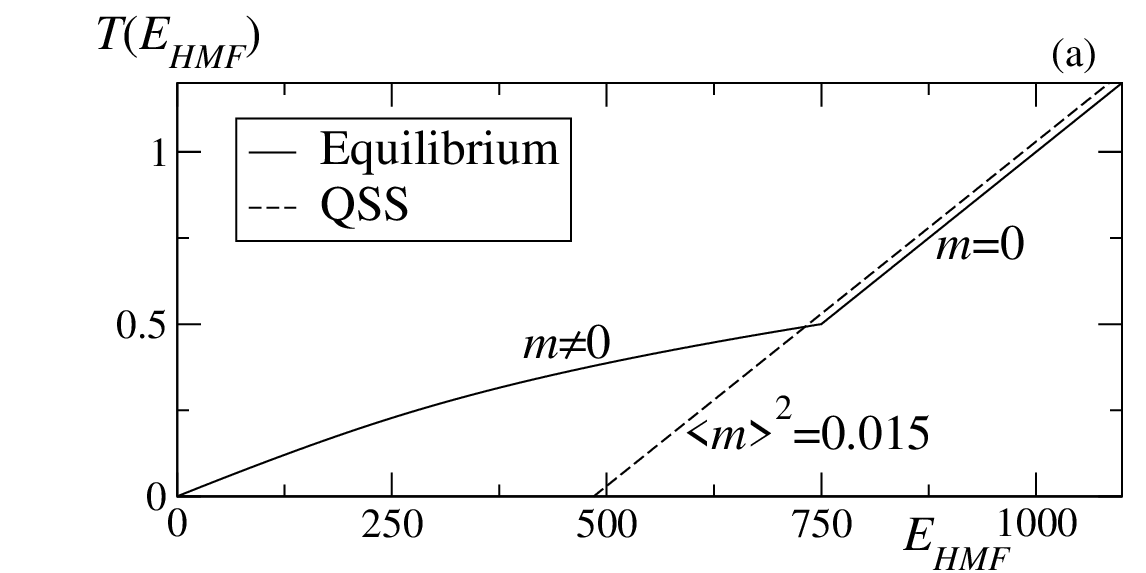}\\
\vspace{0.25cm}
\includegraphics[width=0.98\columnwidth]{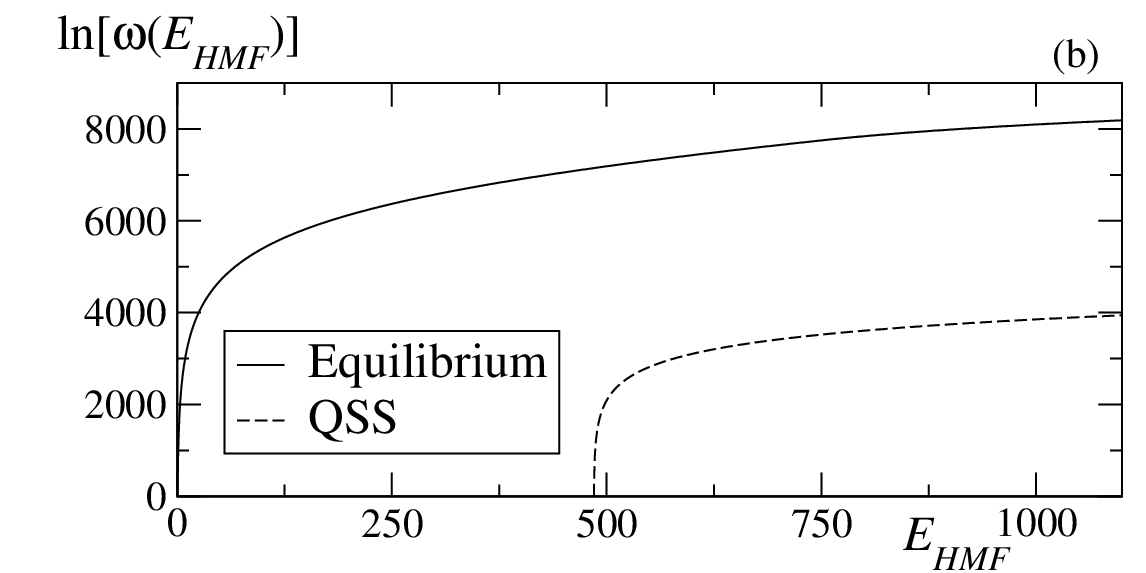}
\caption{
(a): Caloric curve of the HMF model for $M=10^3$. 
Solid line is the BG
equilibrium and dashed line is the curve at fixed $m^2=0.015$. 
(b) Theoretical calculation of $\ln[\omega(E_{\rm HMF})]$ by using
Eq. (\ref{states_density}) with the equilibrium caloric curve (solid line) and with
the curve at $m^2=0.015$ (dashed line).
}
\label{fig_omega}
\end{figure}

\begin{figure}
\includegraphics[width=0.49\columnwidth]{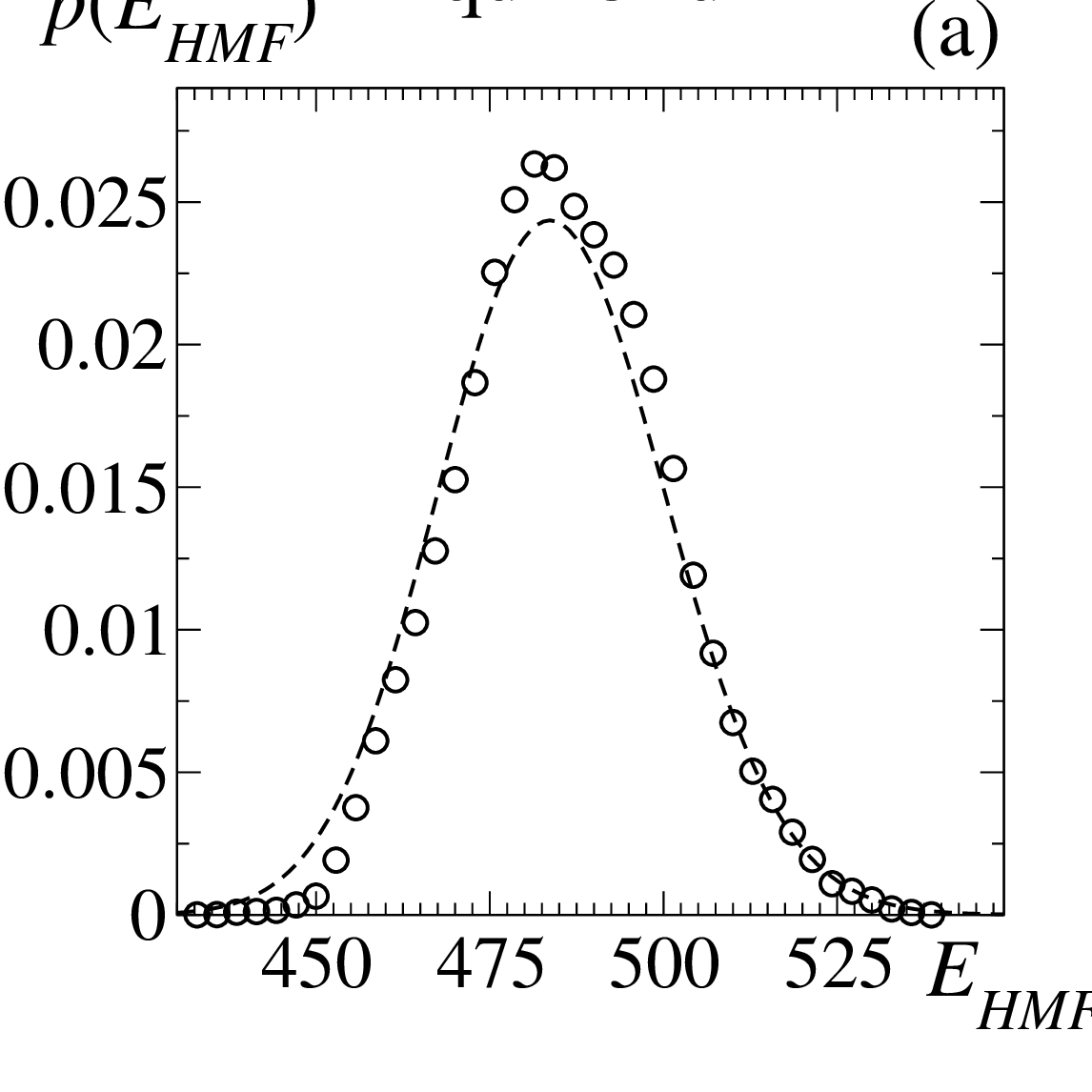}
\includegraphics[width=0.49\columnwidth]{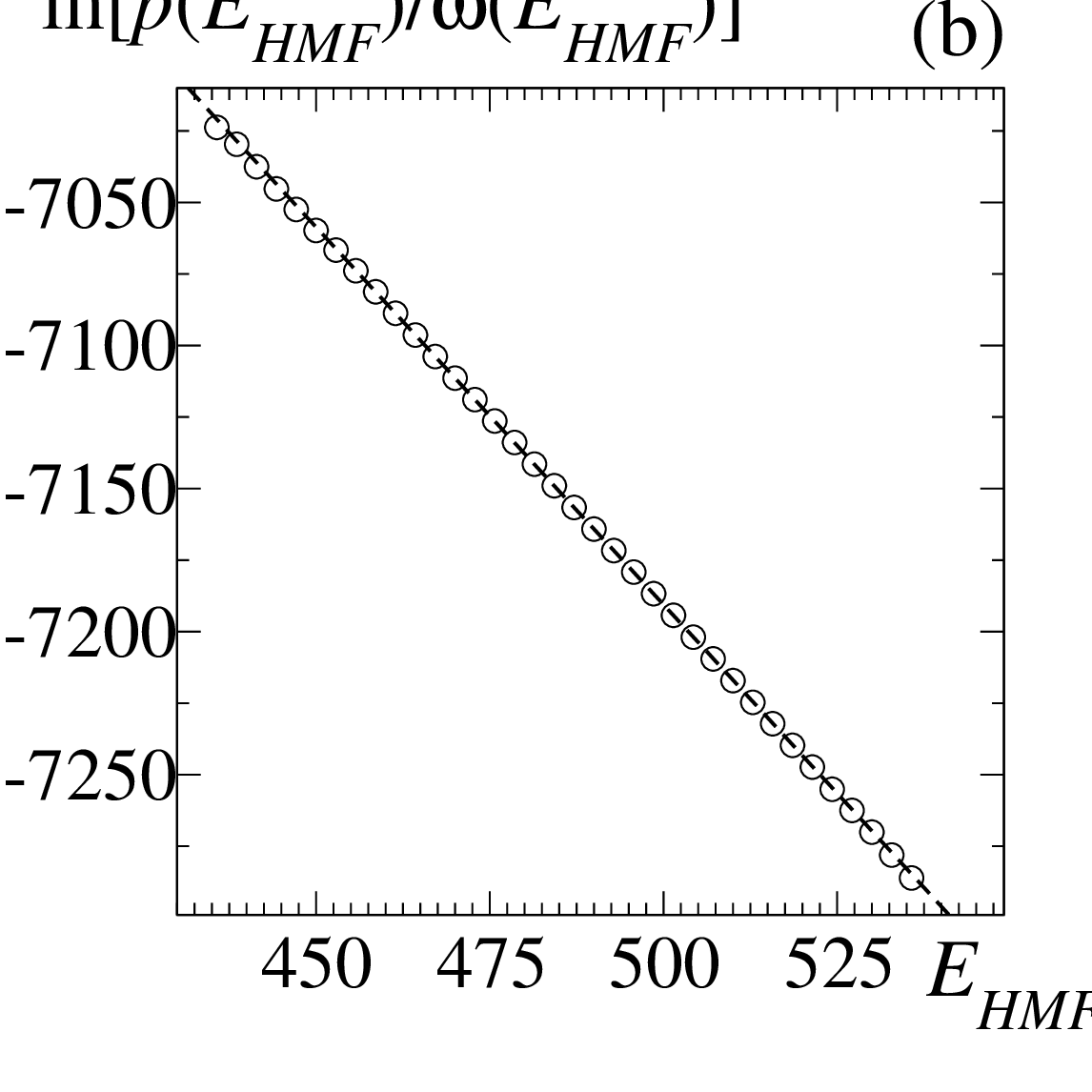}\\
\vspace{0.25cm}
\includegraphics[width=0.49\columnwidth]{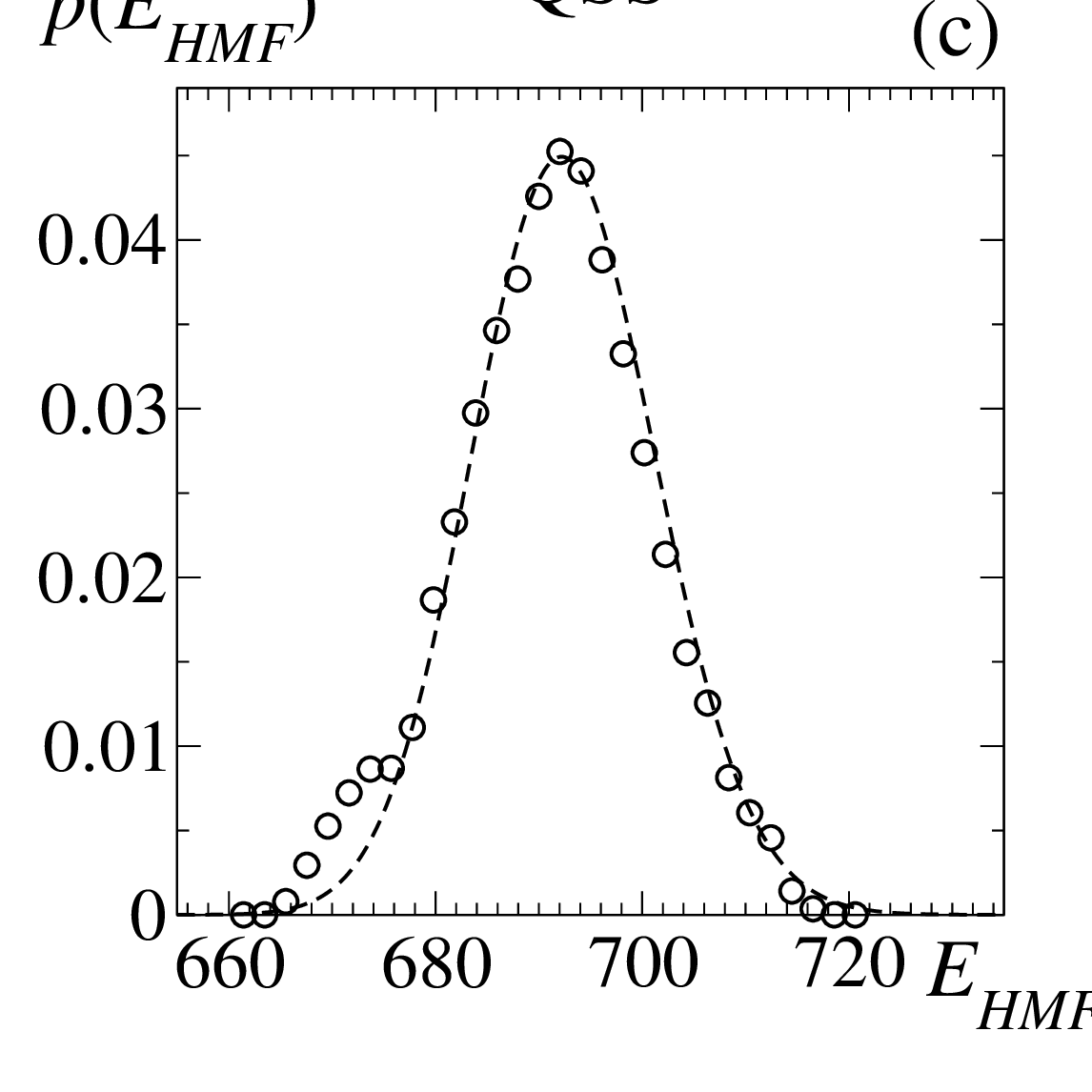}
\includegraphics[width=0.49\columnwidth]{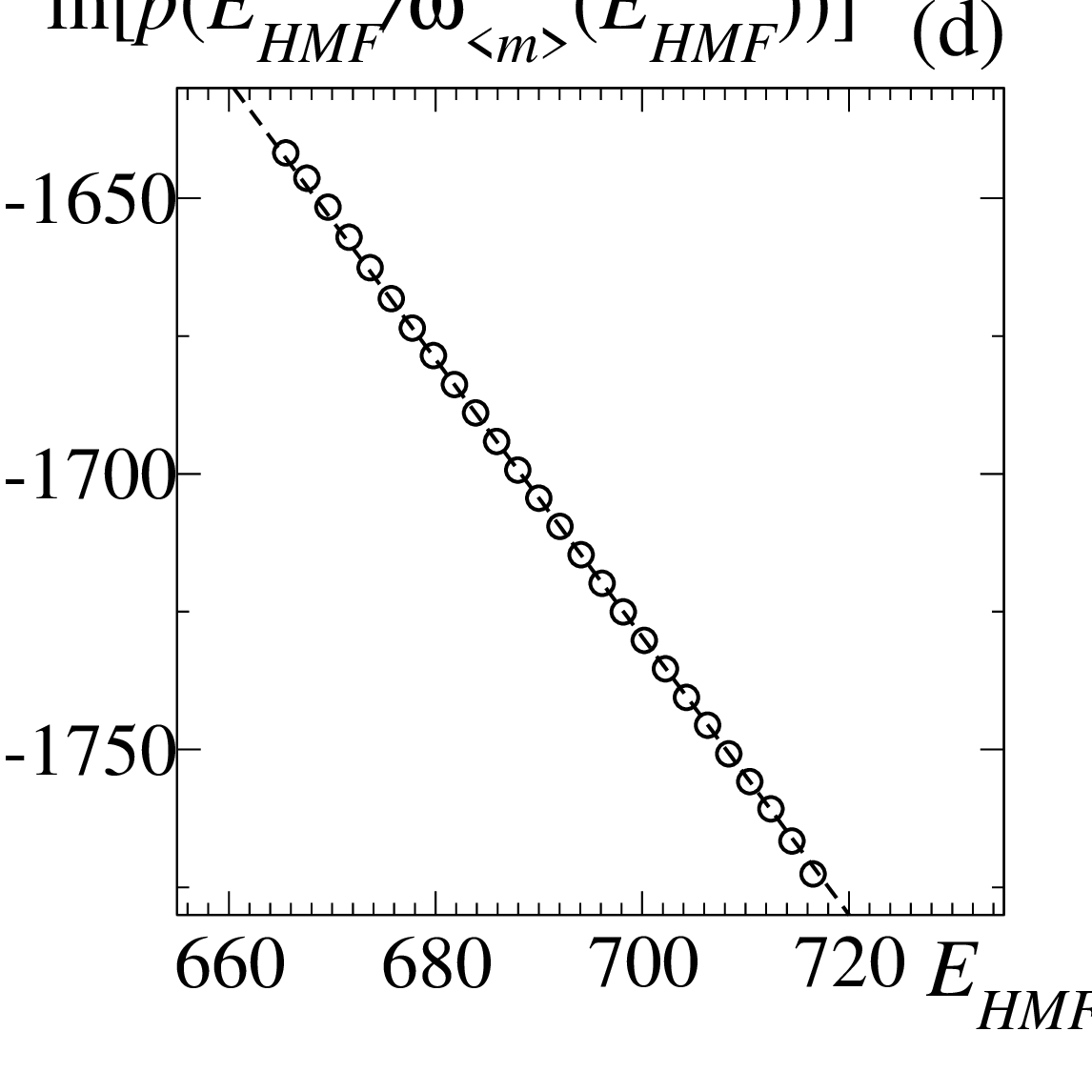}
\caption{
  For $N=10^6$, $M=10^3$ and $\epsilon=0.005$
  comparison between the dynamically recorded $p(E_{\rm HMF})$ (empty circles)
  and $p_{\rm BG}(E_{\rm HMF})$ (dashed lines).  
}
\label{fig_qss_2}
\end{figure}

We now address the main result of the paper, which is central to 
the discussion of the appropriate statistical mechanics approach
for quasi-stationary nonequilibrium states in
long-range systems and to the debate in 
\cite{latora,debate_1,debate_2,debate_3}.
According to BG, the equilibrium PDF of the 
energy $E$ for a system in contact with a TB at temperature $T$ is
$p_{\rm BG}(E)=\omega(E)e^{-E/T}/Z$,
where $Z$ is the partition
function. 
Since the Hamiltonian simulations consent an empirical estimation of
this PDF, it is possible to verify $p_{\rm BG}(E)$ on dynamical basis
\cite{baldovin}. 
From the analytically known solution of the HMF model \cite{konishi,chavanis} one
obtains the BG equilibrium caloric curve of the system $T(E)$  
(full line in Fig. \ref{fig_omega}a).
The integration of the 
thermodynamic relation 
$\partial\ln\omega(E)/\partial E=1/T(E)$,
\begin{equation}
\ln[\omega(E)]-\ln[\omega(E_0)]=
\int_{E_0}^{E} d E'\;\frac{1}{T(E')},
\label{states_density}
\end{equation}
furnishes an analytical
evaluation of $\omega(E)$ (full line in Fig. \ref{fig_omega}b) 
and hence of $p_{\rm BG}(E)$ \cite{baldovin}. 
In Fig. \ref{fig_qss_2}a we show that, as expected, $p_{\rm BG}(E_{\rm HMF})$
and the result of the simulations {\it at equilibrium}, $p(E_{\rm HMF})$,
do coincide.   
A linear regression of $\ln[p(E_{\rm HMF})/\omega(E_{\rm HMF})]$ vs $E_{\rm HMF}$
with a coefficient $R=-0.99997$ gives a direct evidence of the
Boltzmann factor (Fig. \ref{fig_qss_2}b). 
Moreover, the inverse of the slope coefficient
agrees with the dynamical $T=2k_{\rm HMF}$ within an error 
$\Delta T/T=0.3\%$.   

With respect to the QSS, it is interesting to ask 
\cite{latora,debate_1,debate_2,debate_3} if there
exist a statistical mechanics approach that, equivalently to the BG
equilibrium one, can reproduce the dynamically observed $p(E_{\rm HMF})$.  
We first notice that the anomalous dynamical behavior during the QSS
is due to the fact that the system, instead of exploring the
overwhelming majority of $\Gamma$-space microstates, is trapped
\cite{mackay}  in regions characterized by almost constant
nonequilibrium values of the order parameter $m$.
Let $\langle m\rangle$ be the average value around which $m$
fluctuates and $\omega_{\langle m\rangle}(E)$ the sub-manifold of
$\Gamma$-space which corresponds to this dynamical behavior.
The assumption of weak correlations among particles, consistent with
the previous argument based on the CLT and with the Vlasov and
Balescu-Lenard kinetic pictures \cite{balescu}, 
suggests that the Lebesgue measure of
$\omega_{\langle m\rangle}(E)$ is non-zero.
We then expect \mbox{$p(E)=p_{{\rm BG},\langle m\rangle}(E)
\equiv\omega_{\langle m\rangle}(E) e^{-E/T}/Z$} \cite{kinchin}.
Having assumed this, a saddle point calculation 
at fixed $m=\langle m\rangle$ 
(large deviation formulation of the canonical ensemble \cite{ellis} at 
$m=\langle m\rangle$) implies that $T$ in the previous expression
satisfies the fundamental
thermodynamic relation  
$\partial\ln\omega_{\langle m\rangle}(E)/
\partial E|_{E=\langle E\rangle}=1/T$, 
where $\langle E\rangle$ is the average value of the energy during the
QSS. 
Hence, $\omega_{\langle m\rangle}(E)$ can be calculated
by replacing the equilibrium caloric curve $T(E)$ 
with the caloric curve at constant $m=\langle m\rangle$,
$T_{\langle m\rangle}(E)$, and by performing the approach of
Eq. (\ref{states_density}). 
The validity of this strategy, and in particular of Eq. (4) for the
QSS, is further established by the comparison with the dynamical
simulations results. 
Specifically, we show below that 
$T$ corresponds to twice the specific kinetic energy
of the HMF. 

The HMF caloric curve at fixed $m_{\rm HMF}=\langle
m_{\rm HMF}\rangle$ is given, for all $\langle m_{\rm HMF}\rangle\in[0,1]$, by
the straight line  
\begin{equation}
T_{{\rm HMF},\langle m\rangle}(E_{\rm HMF})=2\frac{E_{\rm HMF}}{M}-(1-\langle
m_{\rm HMF}\rangle^2)
\end{equation} 
(e.g., dashed line in Fig. \ref{fig_omega}a for the QSS described in
Fig. \ref{fig_qss_1}).
The integration of the inverse of $T_{{\rm HMF},\langle m\rangle}$ gives 
$\omega_{\langle m\rangle}(E_{\rm HMF})$  
(dashed line in Fig. \ref{fig_omega}b).
The leading behavior of  
$\ln[\omega_{\langle m\rangle}(E_{\rm HMF})]$ is proportional to $M$. This implies
that only an exponential probability for the microstates can balance
this $M$-dependency, to yield an intensive temperature through the relation 
$\partial\ln\omega_{\langle m\rangle}(E_{\rm HMF})/\partial E_{\rm HMF}$.  
In Fig. \ref{fig_qss_2}c it
is shown that $p(E_{\rm HMF})$ observed during the QSS at constant 
$\langle m^2\rangle\simeq0.015$ and 
$\langle2k_{\rm HMF}\rangle\simeq0.397$ agrees with 
$p_{{\rm BG},\langle m\rangle}(E_{\rm HMF})$.
Again, a linear regression of 
$\ln[p(E_{\rm HMF})/\omega_{\langle m\rangle}(E_{\rm HMF})]$ versus $E_{\rm HMF}$
with a coefficient $R=-0.99997$
confirms the Boltzmann factor for 
the energy PDF during the QSS (Fig. \ref{fig_qss_2}d). 
The inverse of the slope coefficient $T$ concurs 
with $\langle2k_{\rm HMF}\rangle$ within an error $\Delta T/T=0.5\%$.  
We checked that a replacement of the limit $\alpha\to0$ in the exponential Boltzmann factor 
$\lim_{\alpha\to0}(1-\alpha\beta E_{\rm HMF})^{1/\alpha}$ with
a finite $|\alpha|\sim 10^{-3}$ 
is already in complete disagreement with
the observed dynamical fluctuations for $M=10^3$.  
We applied the same procedure for different values of
$M$ and to other stationary
and QSSs stemming from different initial conditions
\cite{progress} obtaining similar agreements between our theoretical
scheme and the dynamical simulations. 

We have studied the statistical mechanics of QSSs emerging in
the Hamiltonian dynamics of the HMF model in contact with a reservoir.  
We have shown that weak interparticle correlations and the CLT implies
\cite{kinchin} that the statistical mechanics in $\Gamma$-space is 
obtained by restricting the BG approach to a 
sub-manifold defined by a nonequilibrium
value of the magnetization $m=\langle m\rangle$ \cite{landau}.  
During the QSS, the HMF does not thermalize with the TB.
The temperature to be used in the Boltzmann factor is fixed by the
fundamental thermodynamic relation applied in this nonequilibrium
situation and corresponds to twice the specific kinetic energy of the
system.  
Our theoretical approach, based on the idea of incomplete equilibrium
\cite{landau}, given the quasi-stationary values of the
order parameter and the temperature, allows one to calculate the other
thermodynamic quantities such as the energy of the system and its
fluctuations (i.e., the specific heat). 
We expect the present approach to be significant
for nonequilibrium systems displaying stationarity or
quasi-stationarity
\cite{incomplete,dauxois,barre,latora,morita,progress} 
concomitantly with a kinetic theory based on the Vlasov or
Balescu-Lenard equations \cite{balescu}.

{\bf Acknowledgments}. 
We thank A.L. Stella, C. Tsallis, H. Touchette and S. Ruffo for useful 
remarks.


\begin{thebibliography}{99}

\bibitem{maes}
See, e.g., C. Maes in {\it Mathematical Statistical Physics}, 
A. Bovier et al. eds  (Elsevier B.V., Amsterdam, 2006).

\bibitem{landau}
L. D. Landau and E. M. Lifshitz, Statistical physics  Part 1, 
Pergamon press (Oxford, 3rd edition,
1980);
O. Penrose and J. L. Lebowitz, J. Stat. Phys. {\bf 3}, 211 (1971).

\bibitem{incomplete}
K. R. Yawn and B. N. Miller
Phys. Rev. E {\bf 68}, 056120 (2003);
F. Ritort in {\it Unifying Concepts in Granular Media and Glasses},
A. Coniglio et al. eds. (Elsevier B.V., Amsterdam, 2004).

\bibitem{balescu}
See, e.g., R. Balescu, {\it Statistical Dynamics} (Imperial College Press,
London, 1997).

\bibitem{dauxois}
See, e.g., T. Dauxois, S. Ruffo, E. Arimondo, and M. Wilkens, 
{\it Dynamics and Thermodynamics of Systems with Long Range
  Interactions},  
Lecture Notes in Physics Vol. 602 (Springer, New York, 2002).

\bibitem{ham_can}
F. Baldovin and E. Orlandini, Phys. Rev. Lett. {\bf 96}, 240602
(2006). 

\bibitem{konishi}
T. Konishi and K. Kaneko, J. Phys. A {\bf 25}, 6283 (1992);
M. Antoni and S. Ruffo, Phys. Rev. E {\bf 52}, 2361 (1995);

\bibitem{latora}
V. Latora, A. Rapisarda, and C. Tsallis Phys. Rev. E {\bf 64}, 056134
(2001).

\bibitem{debate_1}
A. Pluchino and A. Rapisarda, Europhys. News {\bf 6}, 202 (2005) and
references therein; 
See also 
A. Rapisarda, A. Pluchino and C. Tsallis, cond-mat/0601409;
C. Tsallis, A. Rapisarda, V. Latora and F. Baldovin in
Ref. \cite{dauxois}.

\bibitem{debate_2}
Y.Y. Yamaguchi, J. Barr\'e, F. Bouchet, T. Dauxois and S. Ruffo,
Physica A {\bf 337}, 36 (2004);
F. Bouchet and T. Dauxois, Phys. Rev. E {\bf 72}, 045103(R) (2005);
P.H. Chavanis, Physica A {\bf 365}, 102 (2006);
A. Antoniazzi, D. Fanelli, J. Barr\'e, P.H. Chavanis, T. Dauxois and
S. Ruffo, cond-mat/0603813.

\bibitem{debate_3}
T.M. Rocha Filho, A. Figueredo, and M.A. Amato, Phys. Rev. Lett. 
{\bf 95}, 190601 (2005);
M.Y. Choi and J. Choi, Phys. Rev. Lett. {\bf 91}, 124101 (2003).

\bibitem{chavanis}
See, e.g., 
P.H. Chavanis, J. Vatteville, and F. Bouchet, Eur. Phys. J. B 
{\bf 46}, 61 (2005) and references therein. 

\bibitem{barre}
J. Barr\'e, T. Dauxois, G. De Ninno, D. Fanelli and S. Ruffo, 
Phys. Rev. E {\bf 69}, 045501(R) (2004).

\bibitem{tamarit}
F. Tamarit and C. Anteneodo, Phys. Rev. Lett. {\bf  84}, 208 (2000);
A. Campa, A. Giansanti and D. Moroni, Phys. Rev. E {\bf 62} 303
(2000); Physica A {\bf 305}, 137 (2002).

\bibitem{kinchin}
A.I. Kinchin, {\it Mathematical
Foundations of Statistical Mechanics} (Dover, New York,
1960).

\bibitem{baldovin}
F. Baldovin, L.G. Moyano and C. Tsallis,  Eur. Phys. J. B {\bf 52}, 
113 (2006). 

\bibitem{mackay}
R.S. Mackay, J.D. Meiss and I.C. Percival,
Physica D {\bf 13}, 55 (1984);
see also
F. Baldovin, E. Brigatti, C. Tsallis, Phys. Lett. A {\bf 320}, 254
(2004).  

\bibitem{ellis}
R. S. Ellis, {\it Entropy, Large Deviations, and Statistical
Mechanics} (Springer-Verlag, New York, 1985).

\bibitem{progress}
F. Baldovin and E. Orlandini, in preparation.

\bibitem{morita}
H. Morita and K. Kaneko, Europhys. Lett. {\bf 66}, 198 (2004);
Phys. Rev. Lett. {\bf 96}, 050602 (2006). 




\end{thebibliography}
\end{document}